\documentclass[12pt]{article}
\usepackage[dvips]{graphicx}

\title{Ways to solve of the evolution equation in the homogeneous
free molecular  condensation under dynamic conditions}
\author{Victor Kurasov}
\date{  }

\begin{document}

\maketitle

\begin{abstract}
The ways to construct  solution
of the evolution equation in the homogeneous
free molecular  condensation under dynamic conditions
are presented in parametric form.
\end{abstract}

Taking the evolution equation in homogeneous case in the power
approximation \cite{Power} one can come to
$$
y^{(IV)} = A y^m
$$
where
$$
y = \zeta /\zeta_0
$$
is the ratio of supersaturations
($\zeta_0$ is a base for approximation)  and $m$ is the big integer
parameter. Here $A$ is the intensity of droplets formation at peak
or at $\zeta =
\zeta_0$.

This equation is studied with the help of methods from
\cite{Polyanin}.


Integration gives
$$
2 y' y^{'''}  - (y^{''})^2  = \frac{2 A}{m+1}
y^{m+1} + 4/3 C
$$
with the arbitrary constant $C$. The substitution
$$
w = (y^{'})^{3/2}
$$
brings the last equation to
$$
w^{''}_{yy} = (\frac{3A}{2m+2} y^{m+1} + C)
w^{-5/3}
$$

Now we shall determine $C$. The value $y=0$ corresponds to the zero
supersaturation. There (and before) there is no formation of
droplets. It means that here the supersaturation is ideal one. The
ideal supersaturation is considered as allowing the linearization.
Then $y'=0$ $y''=0$, $y'''=0$ at $y=0$.
  Then $C=0$.

Then
$$
w^{''}_{yy} = \frac{3A}{2m+2} y^{m+1} w^{-5/3}
$$
 The last equation is the Emden-Fowler equation.
 The
transformation
$$
\xi = \frac{2(m+1) -5/3 +3}{-5/3-1}
y^{\frac{m+1+2}{-5/3-1}} w
$$
$$
u = y^{\frac{(m+1)+2}{-5/3-1}} (y w^{'}_y +
\frac{(m+1)+2}{-5/3-1} w )
$$
leads to the Abel equation
$$
u u^{'}_{\xi} - u = -
\frac{(m+1+2)(m+1-5/3+1)}{(2(m+1)-5/3+3)^2} \xi +
\frac{3A}{2m+2} (\frac{-5/3-1}{2(m+1)-5/3+3})^2
\xi^{-5/3}
$$

The value of parameter $m$ is the big parameter in the theory of
nucleation. Then one can see the behavior of coefficients
$$
\frac{(m+1+2)(m+1-5/3+1)}{(2(m+1)-5/3+3)^2} \rightarrow m^0 \ \ \
at \ \ \ m \rightarrow \infty
$$
$$
\frac{3A}{2m+2} (\frac{-5/3-1}{2(m+1)-5/3+3})^2 \rightarrow m^{-3} \ \ \
at \ \ \ m \rightarrow \infty
$$

Then the last equation be reduced to
$$
u u^{'}_{\xi} - u =   - \xi + \epsilon \xi^{-5/3}
$$
with small $\epsilon$.  The last equation
evidently can be solved with ordinary
perturbation technique.

In the zero approximation the solution can be written in a
parametric form.
$$
\xi= C \exp(-\int \frac{\tau d\tau}{\tau^2 - \tau +1 } )
$$
$$
u = \tau C \exp(-\int \frac{\tau d\tau}{\tau^2 - \tau +1 } )
$$
or
$$
u = \tau  \xi
$$

Beside the power approximation the exponential approximation for
the rate of nucleation is widely used. Since the power
approximation goes to the exponential one after the limit $m
\rightarrow \infty$, the same constructions can be used for the
evolution equation in the exponential approximation.

Although the ways to construct  solution are
given the current solution can not be effectively used. It
is more convenient to use the solution  of the integral equation
directly in the exponential approximation where it has the universal form
\cite{TMF}.

One can also study the case of nucleation on the surface. When the
 embryous growth occurs in the free molecular regime, one can
 quite analogously come to the following equation
 $$
 y^{'''} = A y^m
 $$
 With the help of substitution
 $$
 v(y) = (y^{'})^2
 $$
 it can be reduced to
$$
v^{''}_{yy} = 2 A y^m v^{-1/2}
$$
This is the equation of the Emden-Fauler type
$$
y^{''} = A x^n y^m
$$
Then the already mentioned transition
$$
\xi = \frac{2n+m+3}{m-1} x^{\frac{n+2}{m-1}} y
$$
$$
u = x^{\frac{n+2}{m-1}} (xy^{'}_z + \frac{n+2}{m-1} y )
$$
leads to the Abel equation
$$
 u u^{'}_{\xi} - u  =
 - \frac{(n+2) (n+m+1) }{(2n+m+3)^2} \xi
 + A (\frac{m-1}{2n+m+3})^2 \xi^m
 $$
 Here $n$ is the initial big parameter $m$ and $m$ here is $-1/2$.
 Then
$$
\frac{(n+2) (n+m+1) }{(2n+m+3)^2 } \rightarrow 1/4
$$
$$
(\frac{m-1}{2n+m+3})^2 \sim n^{-2} \rightarrow 0
$$
 So, approximately
$$
 u u^{'}_{\xi} - u  =
 - \frac{(n+2) (n+m+1) }{(2n+m+3)^2} \xi
 $$
which has been already discussed above.

The same constructions  can be used for the exponential approximation.

Here the generalization for the case of decay of metatsable phase
can be directly made.

One can not regard these constructions as some concrete precise
formulas but only as a way to seek a solution. Even if the solution is
presented it is more convenient to use the initial integral form
and to construct the solution numerically or approximately by
methods discussed in \cite{TMF} and the papers cited there.

\end{document}